\documentclass[fleqn,twoside]{article}
\usepackage{epsfig}
\usepackage{amsbsy}
\usepackage{amsfonts}
\usepackage[headings]{espcrc2}

\readRCS
$Id: espcrc2.tex,v 1.2 2004/02/24 11:22:11 spepping Exp $
\usepackage{graphicx}
\usepackage[figuresright]{rotating}

\newcommand{\AmS}{{\protect\the\textfont2
  A\kern-.1667em\lower.5ex\hbox{M}\kern-.125emS}}

\newcommand{\bs}{\begin{slide}}
\newcommand{\es}{\end{slide}}
\newcommand{\be}{\begin{equation}}
\newcommand{\ee}{\end{equation}}
\newcommand{\bea}{\begin{eqnarray}}
\newcommand{\eea}{\end{eqnarray}}

\newcommand{\la}{\left\langle}
\newcommand{\ra}{\right\rangle}
\newcommand{\lc}{\left[}
\newcommand{\rc}{\right]}
\newcommand{\lp}{\left(}
\newcommand{\rp}{\right)}

\newcommand{\bc}{\begin{center}}
\newcommand{\ec}{\end{center}}

\newcommand{\bi}{\begin{itemize}}
\newcommand{\ei}{\end{itemize}}
\def\epm#1#2{\hbox{${\lower1pt\hbox{$\scriptstyle +~#1$}}
\atop {\raise1pt\hbox{$\scriptstyle -~#2$}}$}}
\newcommand{\dat}{\mathrm{dat}}
\newcommand{\art}{\mathrm{art}}
\newcommand{\net}{\mathrm{net}}
\newcommand{\rep}{\mathrm{rep}}
\newcommand{\rmexp}{\mathrm{exp}}

\title{Neural network approach to parton distributions fitting}

\author{The NNPDF Collaboration: Andrea Piccione\address{Dipartimento di Fisica Teorica, 
Universit\`a di Torino, and INFN, Sezione di Torino,\\Via P. Giuria 1, I-10125, Italy},
        Luigi Del Debbio\address{Theory Division, CERN, \\CH-1211 Gen\`eve 23, Switzerland},
        Stefano Forte\address{Dipartimento di Fisica, 
Universit\`a di Milano and INFN, Sezione di Milano, \\Via Celoria 16, I-20133, Italy},
        Jos\'e I. Latorre\address[BCN]{Departament d'Estructura i Constituents de la 
Mat\`eria, Universitat de Barcelona, 
\\Diagonal 647, E-08028 Barcelona, SPAIN} and Joan Rojo\addressmark.}
       
\runtitle{Neural network approach to parton distributions fitting}
\runauthor{Andrea Piccione}

\begin{document}

\begin{abstract}
We will show an application of neural networks to extract
information on the structure of hadrons. A Monte Carlo over experimental
data is performed to correctly reproduce  data errors and correlations. 
A neural network is then trained on each Monte Carlo replica via a genetic
algorithm. Results on the proton and deuteron structure functions, and on
the nonsinglet parton distribution will be shown.
\vspace{1pc}
\end{abstract}

\maketitle

\section{Introduction}

The requirements of precision physics at hadron colliders have
recently led to a rapid improvement in the techniques for the
determination of the structure of the nucleon. Playing this game
factorization is a crucial issue. Indeed, it ensures that we can
extract the parton structure of the nucleon from a process with only
one initial proton (say, Deep Inelastic Scattering at HERA), and then we
can use this as an input for a process where two initial protons are
involved (Drell-Yan at LHC). In the QCD improved parton model the DIS
structure function of the nucleon can be written as
\bea
\label{qcdf2}
F_2 (x,\,Q^2) &=&
x \lc
\sum_{q=1}^{n_f} e_q^2 \,{\cal C}^q\otimes q_q (x, Q^2)  
\right. \\ \nonumber 
&+&  \left. 2 n_f \, {\cal C}^g \otimes g (x, Q^2)\rc\,  
\eea
where $Q^2 = -q^2=-(k-k')^2$, $x = Q^2/2p\cdot q$, and $p$, $k$ and
$k'$ are the momenta of the initial nucleon, the incoming lepton, and
the scattered lepton respectively; ${\cal C}^i$ are the coefficient
functions pertubatively calculable, $q_q(x, Q^2)$ and $g(x, Q^2)$ the
quarks and the gluon distributions that describe the non pertubative
dynamics, the so called Parton Distribution Functions (PDFs). 

The extraction of a PDF from experimental data is not trivial, even if
it is a well estabilished task. In order to do that we have to evolve
the PDFs to the scale of data, perform the $x$-convolution, add theoretical
uncertainties (resummation, nuclear corrections, higher twist, heavy
quark thresholds, $\ldots$), and then deconvolute in order to have a
function of $x$ at a common scale $Q^2$.

Recently it has been pointed out that the uncertainty associated with
a PDFs set is crucial
\cite{Djouadi:2003jg,Frixione:2004us,Tung:2004md}. The uncertainty on
a PDF is given by the probability density $\mathcal{P}\lc f\rc$ in the
space of functions $f(x)$, that is the measure we use to perform the
functional integral that gives us the expectation value
\be
\la \mathcal{F}\lc f(x)\rc\ra=\int\mathcal{D}f
\mathcal{F}\lc f(x)\rc\mathcal{P}\lc f(x)\rc\,,
\ee
where $\mathcal{F}\lc f\rc$ is an arbitrary function of $f(x)$.
Thus, when we extract a PDF we want to determine an
infinite-dimensional object (a function) from finite set of data
points, and this is a mathematically ill-posed problem.

The standard approach is to choose a simple functional form with
enough free parameters ($q(x,Q_0^2)=x^{\alpha}(1-x)^{\beta}P(x)$),
and to fit parameters by minimizing $\chi^2$.  Some difficulties
arise:
errors and correlations of parameters require at least fully correlated 
analysis of data errors;
error propagation to observables is difficult: many
observables are nonlinear/nonlocal functional of parameters;
theoretical bias due to choice of parametrization
is difficult to assess (effects can be large if data are not precise or
hardly compatible).

Here we present an alternative approach to this problem.
First we will show our technique applied to the
determination of the Structure Functions. 
This is the easiest case, since no evolution is required, 
but only data fitting, thus it is a good application to test the technique. 
Then, we will show how this approach can be extended for the
determination of the PDFs.

\section{Structure functions}

The strategy presented in \cite{Forte:2002fg,DelDebbio:2004qj} to
address the problem of parametrizing deep inelastic structure
functions $F(x,Q^2)$ is a combination of two techniques: a Monte Carlo
sampling of the experimental data and a neural network training on
each data replica.

The Monte Carlo sampling of experimental data is performed 
generating $N_{\rep}$ replicas of the original $N_{\dat}$ experimental data,
\bea
F_i^{(\art)(k)} &=&\lp 1+r_N^{(k)}\sigma_N\rp\Big[ F_i^{(\rmexp)}+r_i^{s,(k)}
\sigma^{stat}_i \nonumber \\
&+&  \sum_{l=1}^{N_{sys}}r^{l,(k)}\sigma_i^{sys,l}\Big]\,,  
\eea
where $i=1,\ldots,N_{\dat}$, $\,r$ are gaussian random numbers with the
same correlation as the respective uncertainties, and
$\sigma^{stat},\sigma^{sys}, \sigma_{N}$ are the statistical,
systematic and normalization errors. The number of replicas $N_{\rep}$
has to be large enough so that the replica sample reproduces central
values, errors and correlations of the experimental data.

\begin{figure}[htb]
\begin{center}
\includegraphics[scale=0.4]{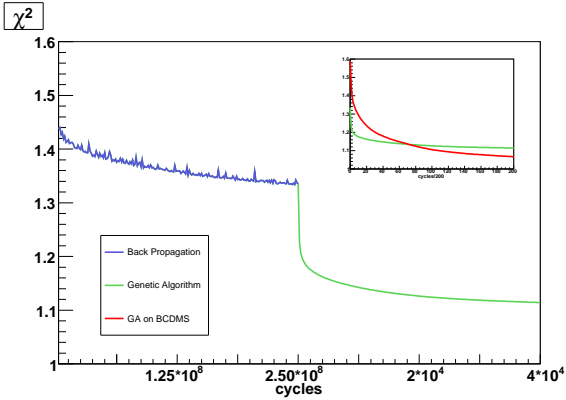}
\caption{Dependence of the $\chi^2$ on the length of training: (big pad) 
total training (small pad), detail of the GA training.}
\label{chi2}
\end{center}
\end{figure}

\begin{figure}[htb]
\begin{center}
\includegraphics[scale=0.4]{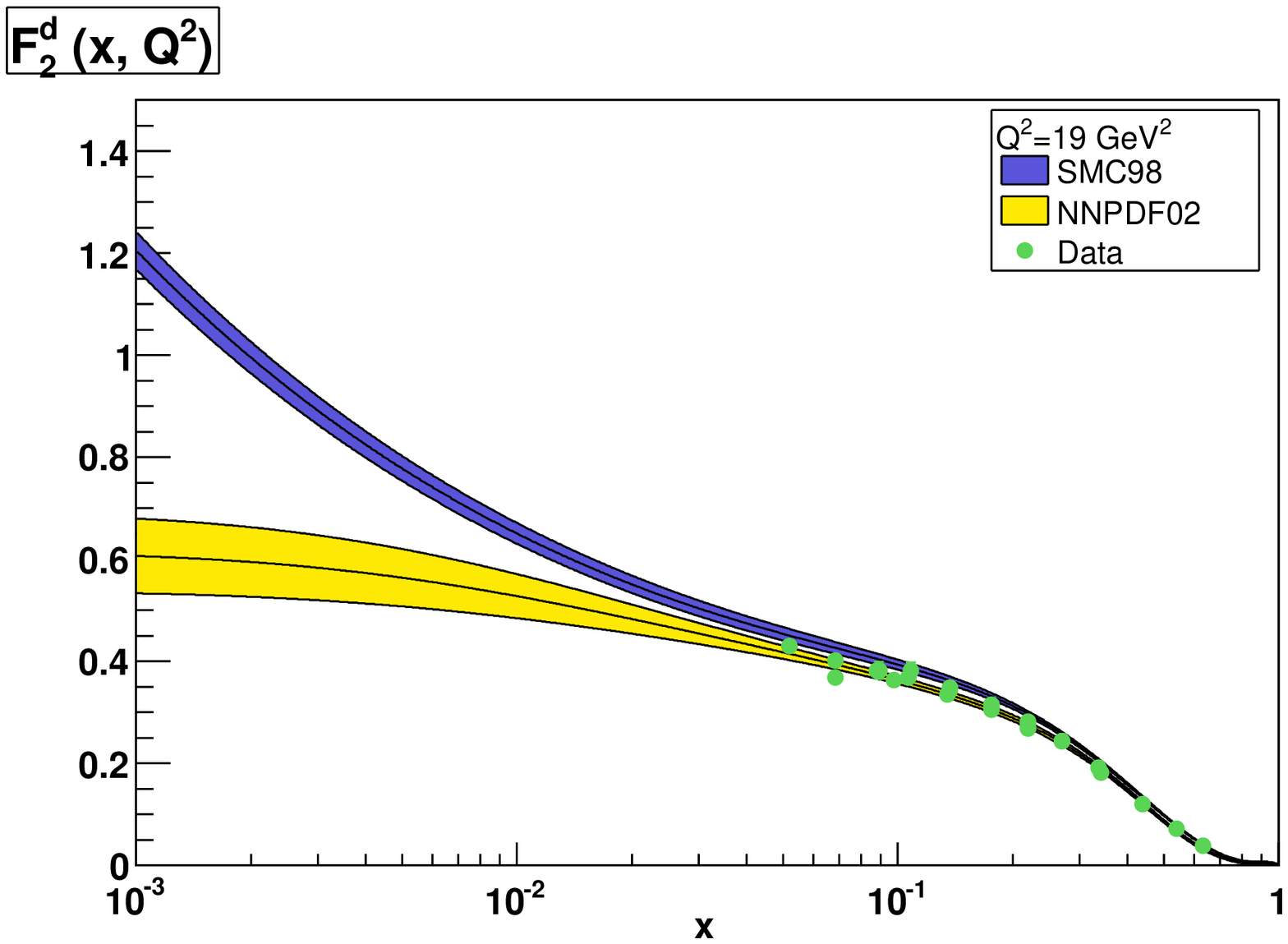}
\includegraphics[scale=0.4]{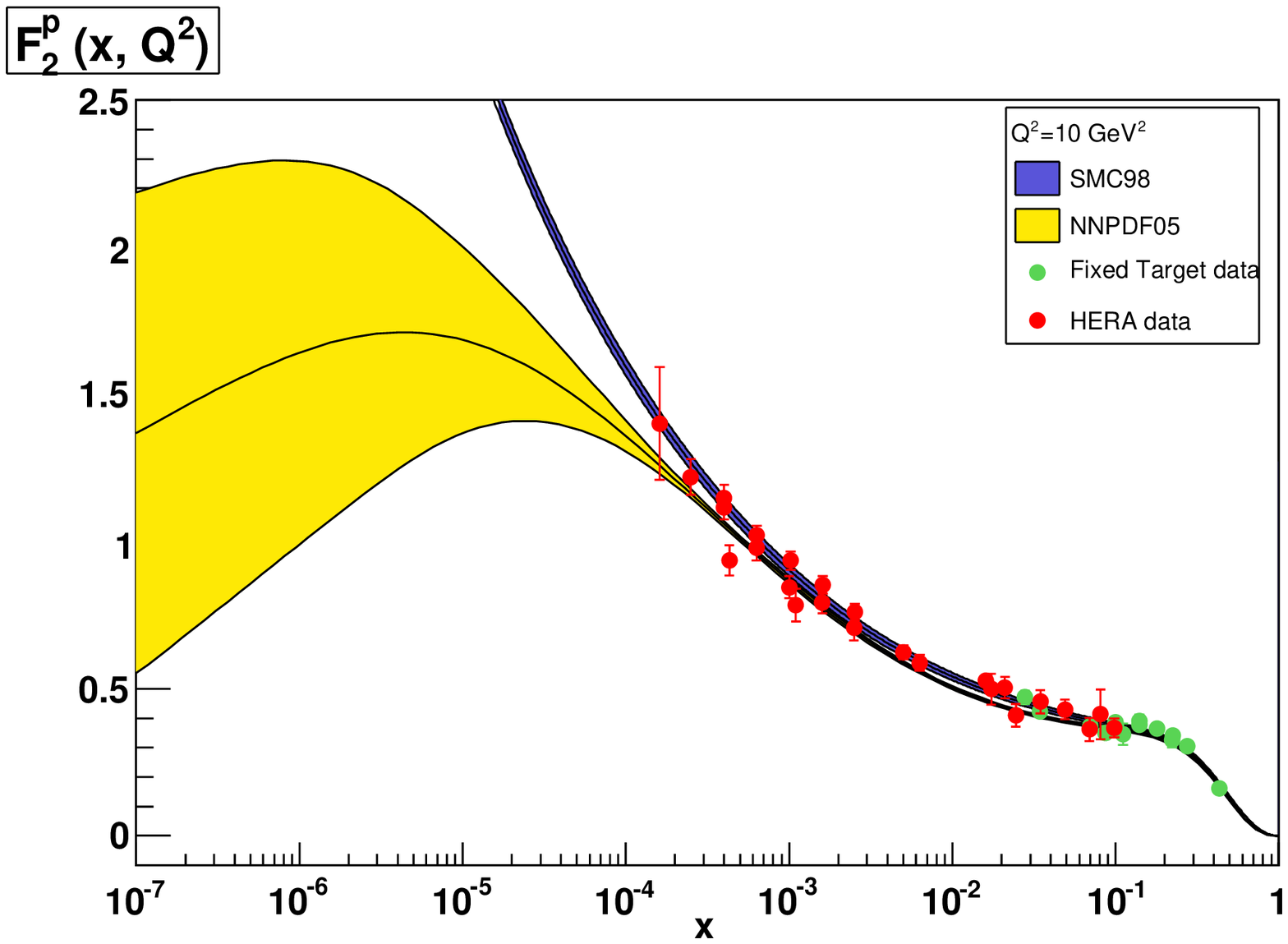}
\caption{Neural networks fit (NNPDF) compared to a polynomial
fit (SMC) for the deuteron and the proton structure function.}
\label{f2dp}
\end{center}
\end{figure}

The second step is to train a neural network on each
data replica. A neural network \cite{Peterson:1991wf} is a highly nonlinear
mapping between input and output patterns as a function of its
parameters. We choose an architecture with 4 inputs 
($x$, $\log x$, $Q^2$, $\log Q^2$),
two hidden layers with 5 and 3 neurons respectively, and one output,
$F(x,Q^2)$. The training on each replica is performed in two steps.
First, we use the Back Propagation technique to minimize 
\bea
{\chi^{2\,(k)}_{\rm diag}}=\frac{1}{N_{\dat}}
\sum_{i=1}^{N_{\dat}}\frac{\lp F_{i}^{(\art)(k)}-
 F_{i}^{(\net)(k)}\rp^2}{(\sigma_{i}^{\it stat})^2}\,;
\eea
then, we use the Genetic Algorithm \cite{Rojo:2004iq} to minimize 
\bea
\chi^{2\,(k)}&=&\frac{1}{N_{\dat}}\sum_{i,j=1}^{N_{\dat}}\lp
F_i^{(\art)(k)}-F_i^{(\net)(k)}\rp
\nonumber \\
&\times & \mathrm{cov}^{-1}_{ij}\lp
F_j^{(\art)(k)}-F_j^{(\net)(k)}\rp\,. 
\label{cme}
\eea
The Back Propagation technique allows for a fast minimization, but
it always oscillates, while the Genetic Algorithm is always
decreasing, and it is more suitable for the last part of the
training where the stability of the $\chi^2$ is needed, see
Fig.~\ref{chi2}. 
Once we have trained all the neural networks on data replicas, we have
a probability density in the space of structure
functions, $\mathcal{P}\lc F(x,Q^2)\rc$, which
contains all information from experimental
data, including correlations. Expectation values over this
probability measure are then evaluated as averages over the trained
network sample,
\be
\label{probmeas}
\la \mathcal{F}\lc F(x,Q^2)\rc\ra=
\frac{\sum_{k=1}^{N_{\rep}}\mathcal{F}\lp F^{(\net)(k)}(x,Q^2)\rp}{N_{\rep}}
\,.
\ee
In Fig. \ref{f2dp} we show our results\footnote{ The source code,
driver program and graphical web interface for our structure function
fits is available at {\tt http://sophia.ecm.ub.es/f2neural}.} for the
deuteron structure function $F_2^d(x,Q^2)$ \cite{Forte:2002fg}, and
for the proton structure function $F_2^p(x,Q^2)$
\cite{DelDebbio:2004qj} compared to a polynomial parametrization \cite{Adeva:1998vv}.
We observe that in the data range the two fits agree within errors.
In the extrapolation region the error band of the polynomial fit has
the same narrow width as in the data range, while the error band of
the neural networks grows indicating that we are in a region where the
error is underterminate since there are no data.

\begin{figure}[htb]
\begin{center}
\includegraphics[scale=0.4]{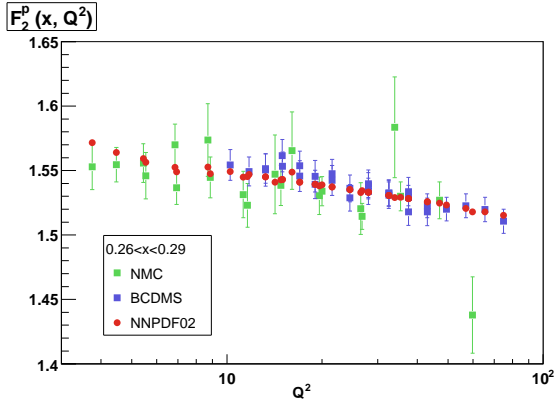}
\caption{Fixed target proton data and predictions for an $x$-bin.}
\label{zoom}
\end{center}
\end{figure}

Neural networks turn to be a suitable tool also in the presence of
uncompatible data. Indeed, once a good fit is obtained, say a stable
value of $\chi^2\sim 1$, the neural networks infer a natural law
by following the regularity of data, and uncompatible data are
discarded without any hypotesis on the shape of the parametrization
(see Fig. \ref{zoom}).

\section{Parton distributions}
The strategy presented in the above section can be used to parametrize
parton distributions as well, provided one now takes into 
account Altarelli-Parisi QCD evolution. 

Now neural networks are used to parametrize the PDF at a reference
scale.  We choose an architecture with 2 inputs ($x$, $\log x$), two
hidden layers with 2 neurons each, and one output,
$q(x,Q_0^2)$. The training on each replica is performed only with
the Genetic Algorithm, since we have a non local function to be
minimized (see eqs. \ref{qcdf2} and \ref{cme}).

\begin{figure}[htb]
\begin{center}
\includegraphics[scale=0.4]{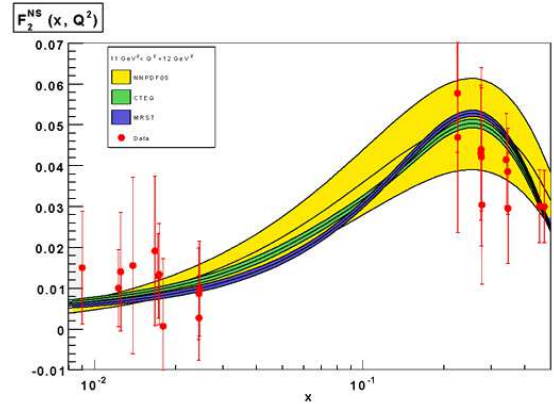}
\caption{Comparison of the prediction for $F_2^{NS}(x,Q^2)$
with different PDF sets.}
\label{qns}
\end{center}
\end{figure}

Once the fit is done, the expectation value and the error
of an arbitrary function $\mathcal{F}$ of a PDF, or
the correlation between different PDFs can be computed
in the following way:
\bea
\la \mathcal{F}\lc q(x)\rc\ra=
\frac{\sum_{k=1}^{N_{\rm rep}}\mathcal{F} \lp q^{({\rm net})(k)}(x) \rp}{N_{\rm rep}}
\nonumber
\eea
\begin{equation}
\sigma_{\mathcal{F}\lc q(x)\rc}=
\sqrt{\la \mathcal{F}\lc q(x)\rc^2 \ra-\la \mathcal{F}\lc q(x)\rc \ra^2}
\end{equation}
\bea
\la u(x_1)d(x_2) \ra=\frac{\sum_{k=1}^{N_{\rm rep}} u^{({\rm net})(k)}(x_1)d^{(net)(k)}(x_2)}
{N_{\rm rep}}
\nonumber
\eea
As a first application of our method, we extract the nonsinglet parton
distribution $q_{NS}(x,Q^2_0)=\frac{1}{6}\lp
u+\bar{u}-d-\bar{d}\rp(x,Q^2_0)$ from the nonsinglet structure
function $F_2^{NS}(x,Q^2)$ measured by the NMC \cite{Arneodo:1996qe}
and BCDMS \cite{Benvenuti:1989rh,Benvenuti:1989fm} collaborations. 
The very preliminary results of a NLO
fit with fully correlated uncertainties can be seen in Fig. \ref{qns}
(only 25 replicas are used instead of 1000). The initial evolution
scale is $Q_0^2=2 \mathrm{GeV}^2$, and the kinematical cuts in order
to avoid higher twist effects are $Q^2=3\,\mathrm{GeV}^2$ and $W^2=6.25\,
\mathrm{GeV}^2$. Our result is consistent within the error bands with
the results from other global fits \cite{Martin:2002aw,Stump:2003yu}, but in
the small-$x$ range where data are poor, differences become more
sizeable. This effect will be further investigated, however, a
larger number of data in the small-$x$ range for the deuteron will
help in cleaning this picture.

Summarizing, we have described a general technique to parametrize
experimental data in an bias-free way with a faithful
estimation of their uncertainties, which has been successfully applied to
structure functions and that now is being implemented in the context of  
global parton distribution fits.

\end{document}